\documentclass[
    reprint,
    preprintnumbers,
    superscriptaddress,
    nofootinbib,
     amsmath,amssymb,
     aps,
     prd,
    floatfix,
    ]{revtex4-2}

    \usepackage{graphicx}
    \usepackage[usenames,dvipsnames]{xcolor} 
    \usepackage{times, mathptmx} 
    \usepackage[utf8]{inputenc}
    \usepackage{color}
    \usepackage{hyperref}
    \usepackage[normalem]{ulem} 
    \usepackage{physics}

    \usepackage{enumitem}
    \usepackage{comment}
    \usepackage{bm}
    \usepackage{bigints}

\hypersetup{
  colorlinks   = true, 
  urlcolor     = blue, 
  linkcolor    = blue, 
  citecolor   = blue 
}
    \allowdisplaybreaks[1]
    \graphicspath{{figures/}}

\usepackage{float}

\newcommand{\oxford}{Astrophysics, University of Oxford, DWB, Keble Road, Oxford OX1 3RH, United Kingdom}
\newcommand{\qmul}{Geometry, Analysis and Gravitation, School of Mathematical Sciences, Queen Mary University of London,
Mile End Road, London E1 4NS, United Kingdom}

\newcommand{\ii}{\mathrm{i}}

\newcommand{\dtc}{\Delta t_\mathrm{c}}

\begin{document}

\title{Self-interacting scalar dark matter around binary black holes}

\author{Josu C. Aurrekoetxea}
\email{josu.aurrekoetxea@physics.ox.ac.uk}
\affiliation{\oxford}
\author{James Marsden}
\email{james.marsden@physics.ox.ac.uk}
\affiliation{\oxford}
\author{Katy Clough}
\email{k.clough@qmul.ac.uk}
\affiliation{\qmul}
\author{Pedro G. Ferreira}
\email{pedro.ferreira@physics.ox.ac.uk}
\affiliation{\oxford}

\begin{abstract}
Gravitational waves can provide crucial insights about the environments in which black holes live. In this work, we use numerical relativity simulations to study the behaviour of self-interacting scalar (wave-like) dark matter clouds accreting onto isolated and binary black holes. We find that repulsive self-interactions smoothen the ``spike'' of an isolated black hole and saturate the density. Attractive self-interactions enhance the growth and result in more cuspy profiles, but can become unstable and undergo explosions akin to the superradiant bosenova that reduce the local cloud density. We quantify the impact of self-interactions on an equal-mass black hole merger by computing the dephasing of the gravitational-wave signal for a range of couplings. We find that repulsive self-interactions saturate the density of the cloud, thereby reducing the dephasing. For attractive self-interactions, the dephasing may be larger, but if these interactions dominate prior to the merger, the dark matter can undergo bosenova during the inspiral phase, disrupting the cloud and subsequently reducing the dephasing. 
\end{abstract}

\maketitle

\section{Introduction}

In the absence of observations of weakly interacting massive particles (WIMPs) \cite{LZ:2022lsv}, light scalar fields like axions have undergone a resurgence as potential candidates for the dark matter (DM) of the Universe \cite{Schive:2014dra} (see \cite{Hui:2021tkt,Urena-Lopez:2019kud,Niemeyer:2019aqm} for reviews). Such light candidates exhibit distinctive signatures on astrophysical scales, potentially influencing the growth of cosmic structure, the cores of galaxies and the environments of compact objects like neutron stars and black holes (BHs). In the last case, which is the focus of this work, detecting the effects of the environment on the gravitational wave signal from merging sources offers one observational possibility \cite{Barack:2018yly,Barausse:2020rsu,LISAConsortiumWaveformWorkingGroup:2023arg,CanevaSantoro:2023aol,Cardoso:2019rou,Barausse:2014tra,Yunes:2011ws,Kocsis:2011dr,Macedo:2013qea,Cardoso:2020lxx,Bertone:2018krk,AlvesBatista:2021gzc,Zwick:2021dlg,Cardoso:2019upw,Maselli:2021men,Amaro-Seoane:2012lgq,Cardoso:2022whc,Cole:2022yzw,Krolak:1987ofj,Koo:2023gfm,Boskovic:2024fga,Karydas:2024fcn,Tomaselli:2024bdd,Tomaselli:2024dbw,Bromley:2023yfi}.

Around spinning BHs, the density of scalar fields can grow through the superradiant instability -- extracting energy and angular momentum from the BH \cite{1971JETPL..14..180Z,Press:1972zz,Zouros:1979iw,Detweiler:1980uk,Cardoso:2004nk,East:2017ovw} (see \cite{Brito:2015oca} for a review). However, in the non spinning Schwarzschild case these fields can also undergo more mundane processes like accretion, where scalar fields are drawn into the deep gravitational wells of the BH \cite{Clough:2019jpm,Hui:2019aqm,Bamber:2020bpu,Bucciotti:2023bvw,deCesare:2023rmg,Sanchis-Gual:2016jst}. For a minimally coupled field obeying the Klein-Gordon equation, accretion leads to the accumulation of a density profile around the BH, with overdensities akin to the ``spikes'' of heavy particle DM candidates \cite{Gondolo:1999ef,DeLuca:2023laa,Berezhiani:2023vlo,Sadeghian:2013laa,Gnedin:2003rj,Merritt:2006mt,Merritt:2003qk,Shapiro:2022prq}. Unlike in the particle case, if the Compton wavelength of the scalar field is close to the horizon radius of the BH, the wave-like nature of the scalar field manifests. In the case of solar mass or supermassive BHs, this corresponds to scalar field masses $m\approx 10^{-11}\,\mathrm{eV}$ and $m\approx 10^{-17}\,\mathrm{eV}$, respectively.

These differences change the impact of wave-like and particle-like candidates on gravitational waves emitted from binary black holes (BBHs). In the case of equal-mass mergers, previous work has indicated that the motion of the binary disrupts particle DM spikes, and thus has little or no effect on the gravitational wave signal near merger \cite{Merritt:2002jz,Bertone:2005hw,Kavanagh:2018ggo,Cole:2022ucw}. Hence, the focus is on extreme/intermediate mass ratio inspirals (EMRIs/IMRIs), where the smaller secondary object does not disrupt the spike of the more massive primary BH.  For scalar DM, on the other hand, we showed in \cite{Aurrekoetxea:2023jwk, Bamber:2022pbs} (see also \cite{Ficarra:2021qeh,Zhang:2022rex,Choudhary:2020pxy,Yang:2017lpm}) that scalar accretion onto equal-mass BHs creates an overdensity between them, which persists during the binary merger. This leads to a faster orbital decay (and hence dephasing of the gravitational-wave signal), which we found is maximized when the Compton wavelength of the scalar field $\lambda_c=2\pi/\mu$ (where $\mu=m c/\hbar$) is of the order of the binary's orbital separation.

Much of our understanding of scalar field DM is based on the simplest model -- a massive scalar field with potential $\mu^2\phi^2$. This approach mirrors the study of particle DM, where most research has focused on non-interacting, perfectly cold DM. As cosmological and astrophysical data have improved, interest has grown in more complex particle DM models that include multiple species, warm initial conditions, and self-interactions \cite{Bertone:2018krk}. Self-interactions can significantly affect galaxy formation, transforming a cusped density profile into a cored profile, and have recently been proposed as a solution to the final parsec problem in the particle case \cite{Alonso-Alvarez:2024gdz,Fischer:2024dte}. Whilst the impact of interactions will be small at average galactic DM densities, they may play a larger role at higher densities if DM accumulates, e.g. around stars or BHs \cite{Budker:2023sex,Shapiro:2014oha,Berezhiani:2023vlo,Kadota:2023wlm,Alonso-Alvarez:2024gdz,Boudon:2023vzl,Feng:2021qkj,Alvarez:2020fyo,Chia:2022udn,Spieksma:2023vwl,Siemonsen:2022ivj,Cannizzaro:2022xyw,Boskovic:2018lkj,Marsden:2024eef,Khlopov:1985fch,Chen:2023vkq,Chen:2019fsq,Chen:2021lvo, Chen:2022oad,Gan:2023swl,Kim:2024rgf}.  In the case of scalar DM, self-interactions enter as higher order terms in the potential, so a quartic term $\lambda\phi^4$ is a natural extension that preserves the parity symmetry. The coupling $\lambda$ parameterises the strength of the self-interactions, with its sign indicating whether they are repulsive or attractive. This type of coupling provides the leading order term for DM with self-interactions, like those expected in the axion potential. Self-interacting scalar fields have been studied in the context of superradiance around isolated BHs, where it was proposed that attractive interactions could lead to rich and unique phenomenology, in particular \textit{bosenova} explosions in which the cloud is radiated and absorbed into the BH \cite{PhysRevLett.86.4211,Arvanitaki:2010sy,Yoshino:2012kn, Yoshino:2015nsa}. 
However, more recent work has shown that in the scalar case the bosenova is unlikely to be reached due to superradiance alone, as the growth saturates when higher modes are excited \cite{Omiya:2024xlz,Omiya:2022gwu,Omiya:2022mwv, Omiya:2020vji,Baryakhtar:2020gao}. Additional accretion may enhance the rate sufficiently to allow the instability to take hold \cite{Hui:2022sri}, and it may play a role in dynamical capture around stars \cite{Budker:2023sex} or the higher densities of  boson stars \cite{Arakawa:2024lqr,Arakawa:2023gyq,Eby:2024mhd}.
Vector bosons are more promising and it has been shown numerically that vector fields with a Higgs-like mass mechanism can excite stringy bosenova explosions for sufficiently strong couplings \cite{East:2022rsi,East:2022ppo}, and tensors should in principle be even more explosive, although the models suffer from some pathologies \cite{East:2023nsk}.

In this paper we use numerical simulations to study the accretion of self-interacting scalar fields onto BHs. First, we describe our setup and chosen model in Sec. \ref{sec:setup}. In Sec. \ref{sec:bh}, we study the evolution of DM near an isolated BH, which gives us information about the effect self-interactions have on the spike of the primary BH of an EMRI system, and highlights the possibility of achieving a bosenova-like explosion via accretion. In Sec. \ref{sec:binary} we use numerical relativity simulations to study the growth of self-interacting scalar clouds during an equal-mass BH merger, and quantify the impact these clouds have on the BBH by extracting the gravitational-wave signal during the inspiral, merger and ringdown. We summarize our results and comment on the physical relevance of our findings in Sec. \ref{sec:discussion}.

\section{Scalar field dark matter model}\label{sec:setup}

We model DM as a minimally coupled complex scalar field $\Phi$ described by the action
\begin{equation}
    S = \bigintssss \dd^4 x \sqrt{-g}\left(\frac{R}{16\pi G} - \frac{1}{2}\left(\nabla_{\alpha}\Phi\right)^* \left(\nabla^{\alpha}\Phi\right) - V(\Phi)\right),
\end{equation}
In particular, we focus on the impact of self-interactions in the DM field by adding a quartic self-interacting term 
\begin{equation}
    V(\vert\Phi\vert) = \frac{\mu^2}{2}\vert\Phi\vert^2 + \frac{\lambda}{4}\vert\Phi\vert^4\,,\label{eq:quartic_potential}
\end{equation}
where $\lambda$ is a real, dimensionful coupling constant\footnote{The quantity $\lambda$ has units of $M^{-2}$ in geometric units. We denote the dimensionless value as $\bar{\lambda}$, see Sec. \ref{sec:discussion} for more details on the conversion.} that parameterises the strength of the self-interactions.  The sign of $\lambda$ specifies whether the self-interactions are repulsive ($\lambda>0$) or attractive ($\lambda<0$). If $\lambda>0$, the potential becomes steeper for larger values of the field, which suppresses growth in the field. For $\lambda<0$ on the other hand, the potential opens up asymptotically, making it easier to explore larger values. However, purely attractive couplings result in a potential that is unbounded from below for sufficiently large values of the field $\vert\Phi\vert$. To avoid possible issues when studying attractive self-interactions, we instead simulate
\begin{align}\label{eq:full_potential}
        V(\Phi) &= \frac{\mu ^4}{2 \lambda} \left(\exp\left[\frac{\lambda}{\mu^2}\vert\Phi\vert^2\right] - 1\right) \\
        &\approx \frac{\mu^2}{2}\vert\Phi\vert^2 + \frac{\lambda}{4}\vert\Phi\vert^4 + \mathcal{O}(\vert\Phi\vert^6)\,, \nonumber
\end{align}
which approximates the quartic potential near the minimum but reaches a plateau rather than turning over after the maximum. 

In the flat-space, homogeneous limit, and in the absence of self-interactions, modelling the DM with a complex scalar field allows us to obtain a stationary density\footnote{Massive, self-interacting complex scalar fields play a fundamental role in describing the behavior of superfluid (zero viscosity) DM \cite{Berezhiani:2015bqa,Berezhiani:2019pzd,Berezhiani:2023vlo}.}. The Noether current associated with the U(1) symmetry of the system corresponds to particle number conservation, making the DM stable. On the complex plane, where the scalar field is defined, the behaviour is that of an ellipse with semi-axis and orientation set by the initial conditions. We choose homogeneous initial conditions $\Phi = (\mathrm{Re}[\Phi],\, \mathrm{Im}[\Phi])$ such that
\begin{align}\label{eq:ics}
    \Phi_0 = (\phi_0,\, 0)\,, \qquad \partial_t \Phi_0 = (0,\, \mu \phi_0)\,.
\end{align}
Then, although the asymptotic value of the fields oscillates ($\Phi \sim \exp[\ii \mu t]$), the modulus $\vert\Phi(t)\vert$ remains constant throughout its evolution and the scalar field tracks circular orbits on the complex plane. This behaviour leads to a constant asymptotic energy density
$\rho(t)=\rho_0$, with the magnitude determined by the initial conditions
\begin{equation}\label{eq:rho_0}
\rho_0 = \mu^2\phi_0^2 + \frac{\lambda}{4}\phi_0^4\,.
\end{equation}
We choose the values $\phi_0$ and $\lambda$ so that the self-interaction term is initially small ($\mu^2 \gg \lambda \phi_0^2 $) and $\rho_0\approx \mu^2\phi_0^2$.

The dynamics of the scalar field is given by the Klein-Gordon equation
\begin{equation}
\nabla^{\alpha}\nabla_{\alpha}\Phi - \frac{\partial V(\vert\Phi\vert)}{\partial \Phi^*} = 0\,,
\end{equation}
where $\nabla_\mu$ is the covariant derivative operator associated to the  metric $g_{\alpha\beta}$. In the first part of this paper we will focus on solving this equation on a fixed Schwarzschild background, so that the metric $g_{\alpha\beta}$ does not evolve. In the second part, we will use numerical relativity to solve the coupled system of the Einstein-Klein-Gordon equations to evolve both the scalar field and the spacetime metric for the case of an equal-mass binary merger.

\section{Dark matter around an isolated black hole}\label{sec:bh}

During an extreme-mass-ratio inspiral, a stellar-mass BH emits gravitational waves as it orbits around a supermassive BH. Given the separation of scales, the smaller BH can be treated as a perturber of the primary BH's spacetime. If the primary has a DM environment, then as the secondary object moves through it, its geodesic motion will deviate from the vacuum case, and this, combined with effects like dynamical friction and accretion, may leave distinctive imprints in the gravitational-wave signal. These signatures are one of the key targets of the Laser Interferometer Space Antenna \cite{Boudon:2023vzl,Boskovic:2024fga,Cole:2022yzw,Maselli:2021men,Zwick:2021dlg,Macedo:2013qea,Barausse:2014tra,Cardoso:2019rou,Rahman:2023sof}. However, characterizing these signatures requires a robust theoretical understanding of the expected environment. Motivated by this, we study the DM overdensity that forms around an isolated BH for different self-interactions, neglecting the backreaction of the DM onto the spacetime of the BH. Specifically, we use the 3+1 open-source code \textsc{grdzhadzha} \cite{Aurrekoetxea:2023fhl} to solve the Klein Gordon equation on a BH background in Cartesian Kerr–Schild coordinates \cite{Visser:2007fj}
\begin{equation}
   d s^2 = \left(\eta_{\mu\nu}+ \frac{2M}{r} l_\mu l_\nu \right) \dd x^\mu \dd x^\nu \,,
\end{equation}
where $\eta_{\mu\nu} =\mathrm{diag}(-1,1,1,1)$ and $M$ is the mass of the BH.  Here $l_\mu = (1,\, x/r,\, y/r,\, z/r)$ is the ingoing null vector with respect to both $g_{\mu\nu}$ and $\eta_{\mu\nu}$. Asymptotically the spacetime is flat and in the absence of self-interactions the scalar field oscillates with frequency proportional to its mass $\Phi \sim \exp[\ii \mu t]$. Closer to the BH, the field is accreted and its amplitude and hence density grow \cite{Clough:2019jpm,Hui:2019aqm,Bamber:2020bpu}. 
After a short time, it reaches a quasi-stationary state with a density profile that (in the case of $\mu M \sim 1$) is proportional to $r^{-3/2}$. Notably, since our boundary conditions effectively feed in more scalar field over time, in the purely massive case, the overall amplitude of the cloud continues growing and does not stabilize, although the shape of the profile remains the same.

We start from the homogeneous initial conditions in Eqn. \eqref{eq:ics} with $\phi_0=0.2 M$, and study the evolution of the self-interacting scalar field around a BH of mass $M$. We study cases where the mass of the scalar $\mu =M^{-1}$ and $\mu =0.1M^{-1}$ to explore lower and higher masses, although in both cases the Compton wavelength of the scalar field is roughly of the same order of the Schwarzschild radius.  These values are partly set by the numerics but also cover the wave-like regime where the density near the BH is not strongly suppressed by the pressure effects arising from a large Compton wavelength (unlike relatively low mass scalars $\mu M \ll 1$). We use a grid resolution of $N^3 = 128^3$ with $6$ levels of refinement, and a physical box size of $L=576 M$, which gives a grid spacing of $0.07M$ at the BH horizon.

We use diagnostics that follow the approach of \cite{Clough:2021qlv,Croft:2022gks,Traykova:2021dua,Traykova:2023qyv,Aurrekoetxea:2023jwk}, and define a current $J^\mu=\xi^\nu T^\mu_\nu$ in the direction $\xi^\nu$, as well as an associated charge and flux
\begin{equation}
    Q =-n_\mu J^\mu\,, \qquad F= \alpha N_i J^i\,,
\end{equation}
where $N_i$ is the outward normal direction to the surface that bounds the volume. If $\xi^\nu$ is a Killing vector then $\nabla_\mu J^\mu=0$ and the change in charge is balanced by a flux through a surface. For the fixed background case $\xi^\nu_t=(1,0,0,0)$ is a Killing vector, which gives us the density and fluxes across surfaces
\begin{align}
    \rho_E \equiv Q_t &= -\alpha \rho +\beta_k S^k\,, \label{eq:Qt}\\
    F_E \equiv F_t &=  N_i\left(\beta^i(\alpha \rho - \beta^j S_j) + \alpha(\beta^k S_k^i - \alpha S^i\right)\,,\label{eq:Ft}
\end{align}
where $N_i = s_i / \sqrt{(\gamma^{jk} s_j s_k)}$is the normalised radial unit vector, with $s_i=(x,y,z)/r$. Here $\alpha=1/\sqrt{-g^{00}}$ and  $\beta_i=g_{0i}$. The quantities $\rho$ and $S_i$ are the components of the stress-energy tensor measured by the normal observers
\begin{equation}
    \rho = n^\mu n^\nu T_{\mu\nu}\,, \qquad S_i = n^\mu T_{\mu i}\,,
\end{equation}
where $n_\mu=(-\alpha,0,0,0)$.

\begin{figure*}[t!]
    \centering
    \includegraphics[width=\linewidth]{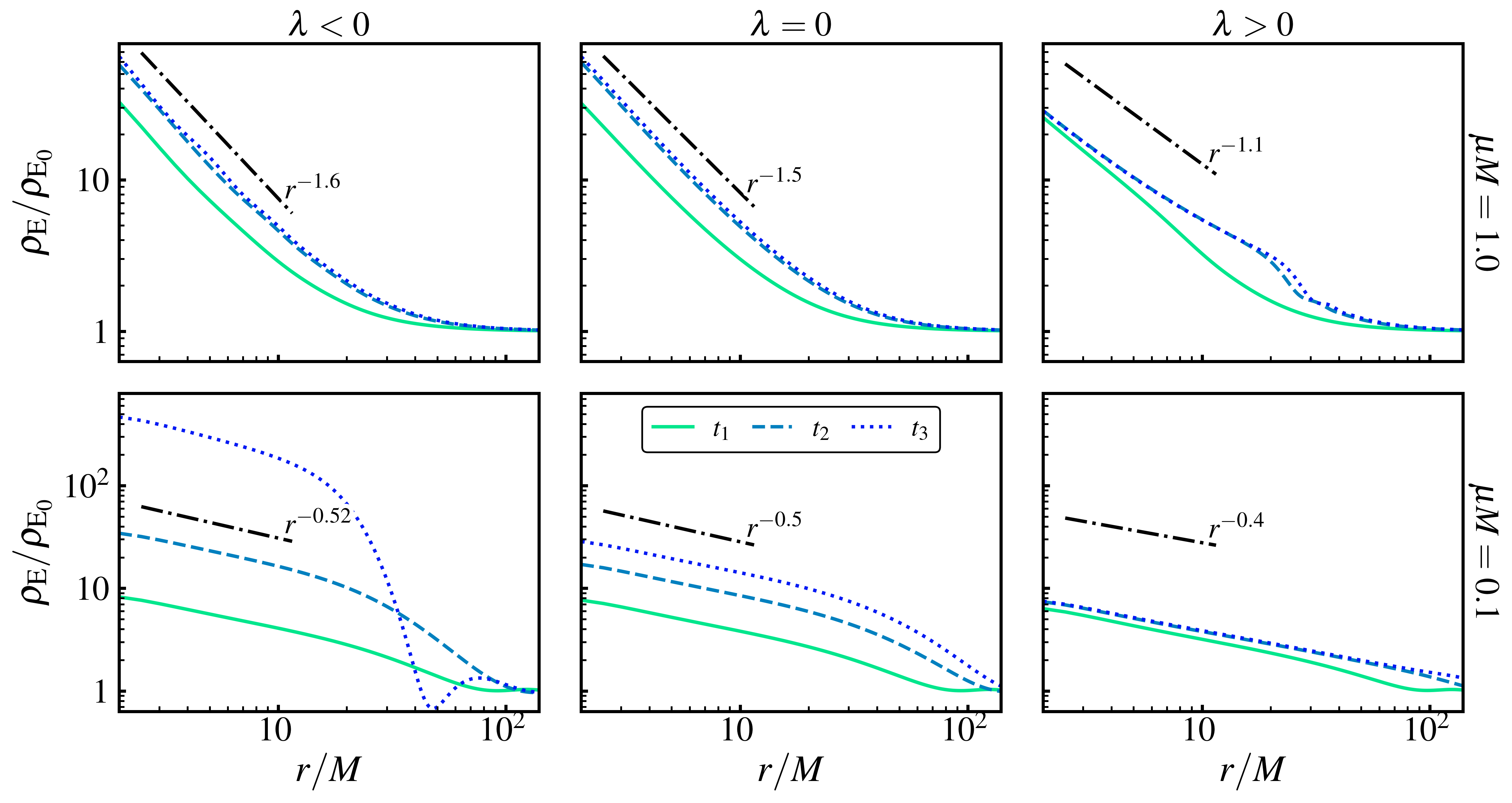} 
    \caption{Growth of the dark matter density profile for different values of the mass $\mu$ and self-interaction $\lambda$. The top (bottom) panel shows the high (low) mass case as labelled. The left panels plot the evolution of the radial profiles for attractive self-interactions $\lambda=-150$ ($\lambda=-0.003$), the middle for the purely massive cases and the right panel for repulsive cases $\lambda=1.6$ ($\lambda=0.015$). The different lines show the field density profiles over time $\lbrace{t_1,t_2,t_3\rbrace} = \lbrace{100,\, 200,\, 220\rbrace} M$ ($\lbrace{t_1,t_2,t_3\rbrace} = \lbrace{200,\, 600,\, 1000\rbrace} M$), normalised by the initial (homogeneous) density. The black dashed line shows a power law fit to the lines. We expect the power law profiles to continue to propagate outwards over time, but are limited in the length of our simulations by the domain size.}\label{fig:rho_profiles_panels}
\end{figure*}

In Fig. \ref{fig:rho_profiles_panels} we plot radial profiles of the density $\rho_E$ at different times. The top and bottom panels correspond to $\mu M=1$ and $\mu M=0.1$ respectively, and each column corresponds to attractive ($\lambda<0$), free ($\lambda=0$) and repulsive ($\lambda>0$) self-interactions. For $\mu M = 1$, the repulsive case ($\lambda>0$) exhibits two interesting properties compared to the density profiles of the free and attractive interaction cases: (i) The profile is flatter towards the centre, resulting in a more ``cored'' structure. (ii) The density saturates, leading the profile to settle into a stable configuration. In contrast, the attractive ($\lambda<0$) and non-self-interacting ($\lambda=0$) cases show minimal differences from each other within the limited time range shown here. This limitation arises because the dynamics of scalar fields in concave potentials are more challenging to resolve numerically when the timescale of the oscillation is short. We also study the low-frequency regime by decreasing the mass of the scalar field to $\mu M = 0.1$. We plot the results in the bottom panel of Fig. \ref{fig:rho_profiles_panels}. In this limit, the power laws are approximately the same, but the results show that attractive self-interactions ($\lambda < 0$) enhance the growth of the cloud, while repulsive self-interactions ($\lambda > 0$) stabilize and saturate the growth. It can be seen that the field density first stabilizes at the horizon, and as accretion continues, it extends out to larger radii. The accretion flux stabilises at a maximum value and the scalar field is able to reach a stationary state.

\begin{figure}[b!]
    \centering\includegraphics[width=0.99\linewidth]{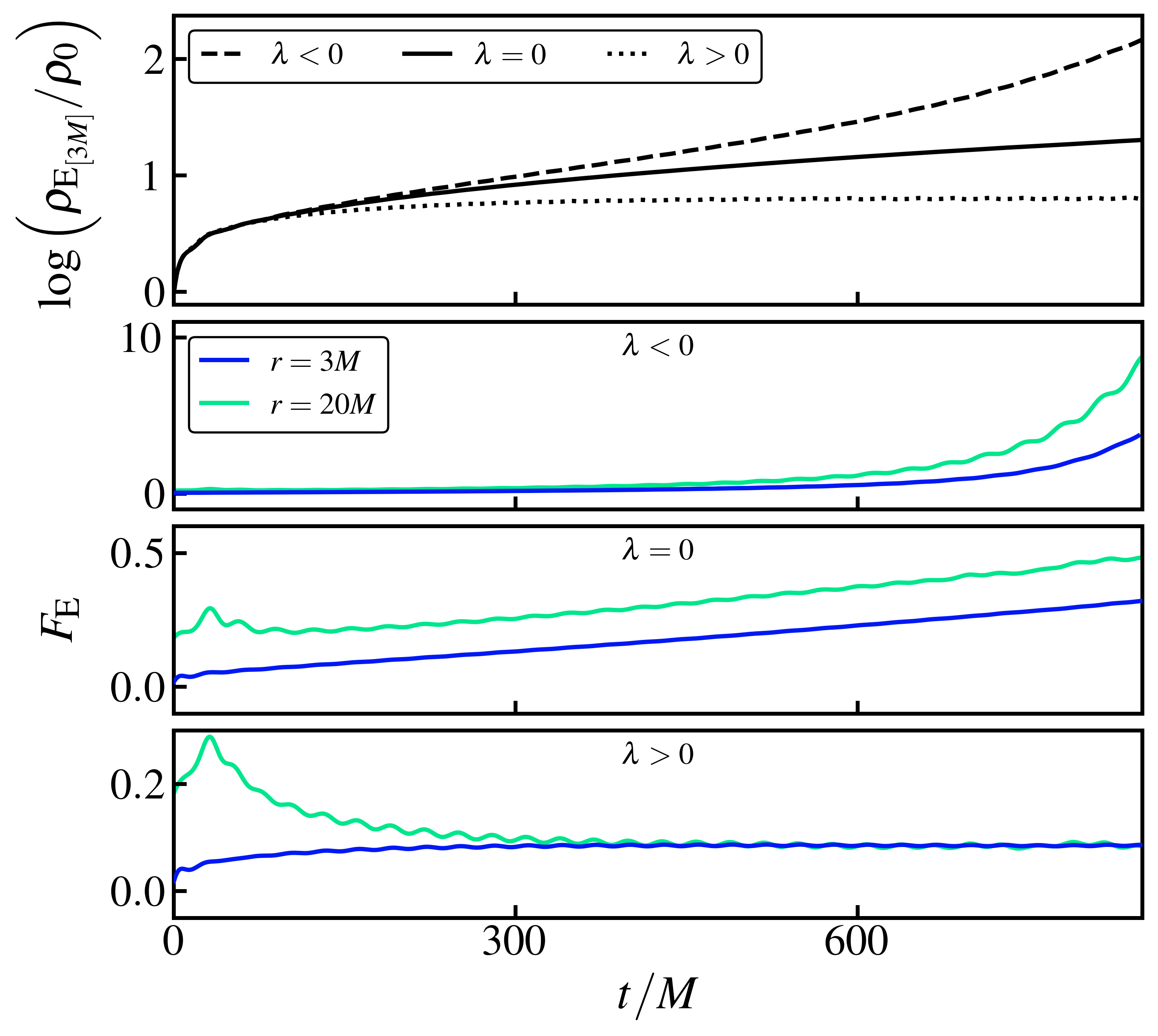}
    \caption{Top panel: Density of the dark matter cloud at $r=3M$ for different values of $\lambda=-0.003$, $\lambda=0$ and $\lambda=0.015$. Bottom panels: Fluxes at radii $r=3M$ and $r=20M$. The growth of the density can be understood from the differences in the fluxes between these radii, which diverge, saturate or reach zero for each of the cases.}
    \label{fig:rho_saturation}
\end{figure}

This saturation and enhancement of the scalar cloud, depending on the sign of the self-interactions, is more apparent in the top panel of Fig. \ref{fig:rho_saturation}, where we plot the evolution of the density at $r=3M$. Initially, the scalar field evolves with small oscillations in the quadratic part of the potential for all three cases, so the growth due to accretion shows negligible differences at early times. However, as the scalar field amplitude increases, it begins to probe the quartic part of the potential, leading to variations in the density profile. The attractive case shows an enhancement in the growth of the cloud compared to the $\lambda=0$ case, while repulsive self-interactions result in the saturation of the cloud's growth. We interpret these results using the bottom three panels of Fig. \ref{fig:rho_saturation}, where we extract the fluxes at two different radii, $r=3M$ and $r=20M$. The difference between the fluxes, $\Delta F$, indicates the buildup of energy within the volume. The panels show that for the attractive case, $\Delta F$ grows faster than in the $\lambda=0$ case, leading to an enhancement in the cloud's density. In contrast, for the repulsive case, the outer and inner fluxes balance perfectly, causing the density profile to saturate and remain constant.

We understand the saturation of the repulsive case as follows. The stability of the scalar cloud is related to the balance of the Newtonian gravitational potential, the gradient pressure of the scalar field, and the pressure induced by the self-interactions $\lambda$. When the scalar field probes the quadratic part of the potential, the gravitational potential dominates and results in the accretion and growth of the cloud. As we have seen, this implies that the scalar field increases its amplitude and starts probing the quartic part of the potential. In the $\lambda>0$ case, the interaction potential creates a repulsive ``pressure'' that allows the scalar cloud to reach hydrostatic equilibrium, with the repulsive self-interaction balancing the self-gravity.

\begin{figure}[t!]
    \centering
    \includegraphics[width=\linewidth]{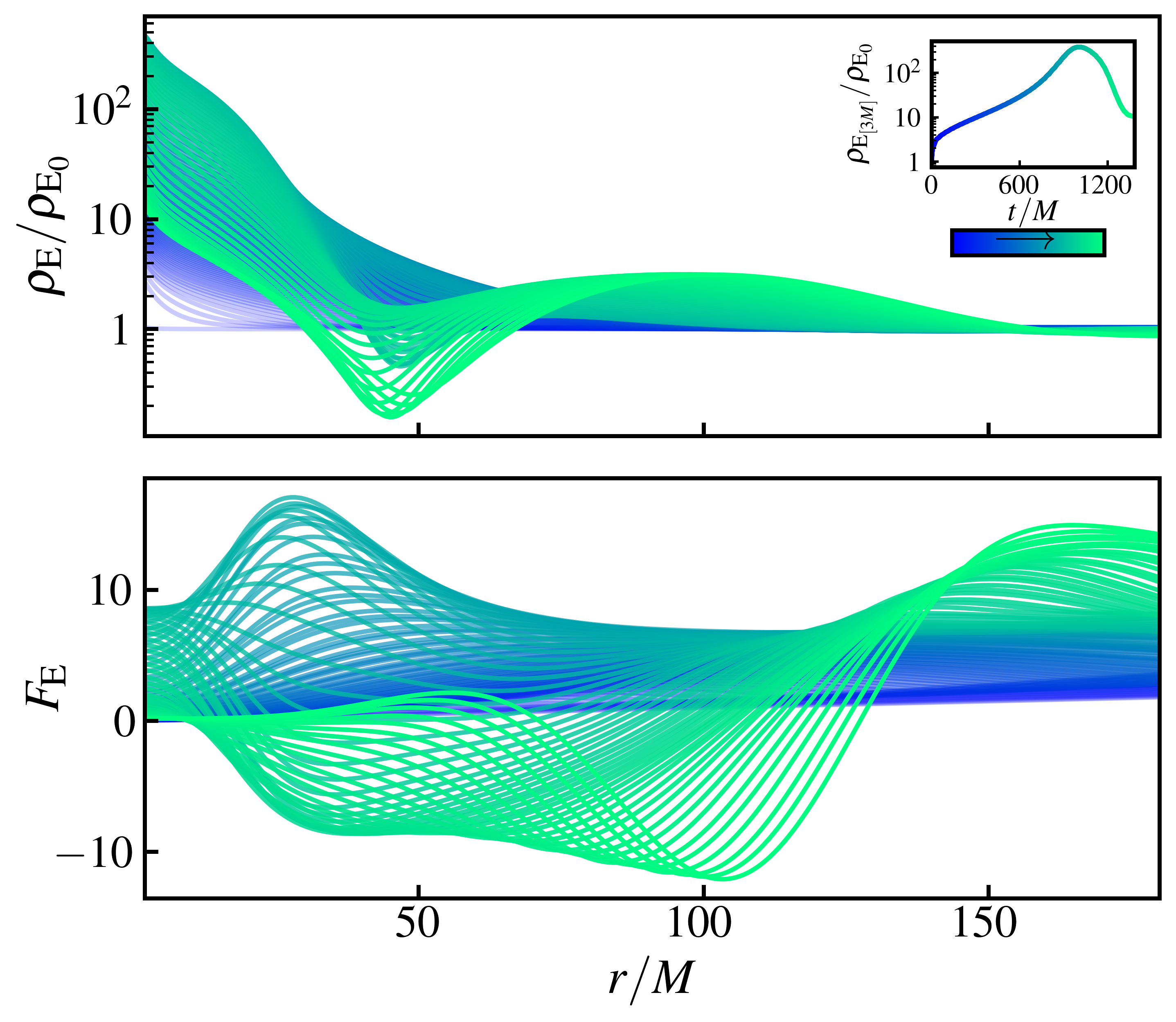}
    \caption{Evolution of the density and fluxes for the bosenova explosion for $\lambda=-0.003$. The top (bottom) panel shows the radial profiles of the density (radial flux). The inset panel tracks the density value at $r=3M$. Positive and negative fluxes correspond to ingoing and outgoing radiation, respectively. The density reaches a peak following which the cloud collapses, and the system undergoes a violent burst of energy that propagates outwards.}
    \label{fig:bosenova}
\end{figure}

The interaction changes sign and becomes attractive when $\lambda<0$, so the only contribution that can lead to hydrostatic equilibrium is the gradient pressure of the scalar field. Attractive self-interactions enhance the growth of the cloud, but the gradient pressure cannot compete with the attractive force due to self-interactions. When the density becomes too high, the system undergoes a rapid collapse -- a bosenova-like event. This collapse increases the density and potential energy, leading to an explosive ejection of energy from the cloud and a subsequent reduction in density. The inset panel of Fig. \ref{fig:bosenova}, showing the evolution of density at $r=3M$, illustrates this process. The evolution of the density profile is shown in the top panel, with blue to green colours (and transparency) depicting time. The density starts homogeneous but grows near the BH. The density increases until $t\approx 1000M$, when it suddenly collapses and the cloud xplodes, radiating part of its mass outwards. In the bottom panel we plot the evolution of the radial fluxes. Initially, the fluxes are positive everywhere, so that there is an ingoing flux of DM towards the BH. The density grows, so the field explores flatter regions of the potential, so that the density and flux is enhanced. Eventually, it reaches a maximum at $r\approx 25M$, after which the flux becomes negative,  resulting in the explosion that reduces the density near the BH. This matter propagates outwards, shown by the the negative fluxes at late times in the bottom panel of the figure.

\section{Dark matter around an equal mass black hole merger}\label{sec:binary}

As discussed in the introduction, unlike in the particle DM case, equal-mass BHs immersed in scalars with Compton wavelength similar to the binary separation may retain the environment up to merger. This motivates the study of equal-mass binaries in such environments including backreaction, to quantify the potential impact on the GW signal. In such a case, the strong-field dynamics of the system needs to be resolved through numerical relativity. 
We decompose the line element in the general ADM form \cite{Arnowitt:1962hi}
\begin{equation}
    ds^2 = -\alpha^2 \dd t^2 + \gamma_{ij}(\dd x^i + \beta^i \dd t)(\dd x^j + \beta^j \dd t)\,.
\end{equation}
Here $\gamma_{ij}$ is the 3-dimensional metric, and the quantities $\alpha$ and $\beta^i$ determine the choice of spatial hyperslicings and their coordinates, which follow the 1+log moving puncture gauge \cite{Bona:1994dr,Baker:2005vv,Campanelli:2005dd,vanMeter:2006vi}. The extrinsic curvature tensor $K_{ij}=(2 D_{(i} \beta_{j)}-\partial_{t} \gamma_{ij})/2\alpha$ is decomposed into a trace $K$ and a traceless part $A_{ij}$, i.e. $K_{ij} = A_{ij}+(1/3)K\gamma_{ij}$. We evolve the coupled system of Klein-Gordon and general relativity equations in the CCZ4 formulation \cite{Alic:2011gg} using the 3+1 open-source numerical relativity code \textsc{grchombo} \cite{Andrade:2021rbd,Radia:2021smk,Clough:2015sqa}.

We study the tensor gravitational-wave modes emitted by the BBH. To do so, we extract the Newman-Penrose scalar $\Psi_4$ with tetrads proposed by \cite{Baker:2001sf}, projected into spin-weight $-2$ spherical harmonics,  
\begin{equation}
\psi_{lm}=\bigointssss_{S^2}\Psi_4{\big\vert}_{r=r_\mathrm{ex}}\left[{}_{-2}\bar{Y}^{lm}\right]\,\dd \Omega\,,
\end{equation}
where $\dd \Omega = \sin\theta\,\dd\theta\,\dd\varphi$ is the area element on the $S^2$ unit sphere.
The merger or coalescence time for the 10-orbit binary in the absence of a dark matter cloud is $\bar{t}_\mathrm{c}\approx 2000\,M$, defined as when the amplitude of the dominant mode $\vert\psi_{22}\vert$ peaks. In Ref. \cite{Aurrekoetxea:2023jwk}, we showed that scalar DM around the binary induces a faster orbital decay, so that the gravitational-wave signal experiences a dephasing/delay of the coalescence time
\begin{equation}
    \dtc \equiv t_\mathrm{c} - \bar{t}_\mathrm{c}\,.
\end{equation}
In particular, we found that the dephasing is maximized when the Compton wavelenght of the scalar particle matches with the initial orbital separation of the binary, $2\pi/\mu\approx d_0$. For our 10-orbit binary, where $d_0\approx 12M$, this corresponds to a scalar particle with mass $\mu\approx 0.5 M^{-1}$.

In this work we investigate the impact that repulsive and attractive self-interactions have in the growth of the DM cloud and the dephasing of the binary. For $\lambda>0$, we evolve the system with the quartic potential in Eqn. \eqref{eq:quartic_potential}. For $\lambda<0$, we simulate the potential in Eqn. \eqref{eq:full_potential} to avoid unbounded values for large amplitudes of the field. We start from homogeneous initial conditions given by Eqn. \eqref{eq:ics} and choose Bowen-York initial data for $A_{ij}$ for the boosted BHs \cite{Bowen:1980yu}. We solve the Hamiltonian and momentum constraints using the CTTK-hybrid method \cite{Aurrekoetxea:2022mpw}, so that the trace of the extrinsic curvature tensor is initially $K_0^2=24\pi G\rho_0$. We fix $\rho_0\approx 10^{-9}M^{-2}$, $\mu=0.43M^{-1}$, and vary both $\lambda$ and $\phi_0$ such that $\rho_0$ is unchanged\footnote{Note that given the small asymptotic densities used in our model, the spacetime is effectively asymptotically flat.} via Eqn. \eqref{eq:rho_0}. This approach ensures that we only need to solve the constraints once and the BHs have identical initial masses, velocities and directions for all our different runs. Then, any observed differences arise solely from their interaction with the DM cloud, and depend on the relative enhancements in the cloud density achieved for different scalar masses and interactions.

\begin{figure}[t!]
    \centering
    \vspace{-20pt}
    \href{https://youtu.be/ZeoMlUHknUM}{
    \includegraphics[width=\linewidth]{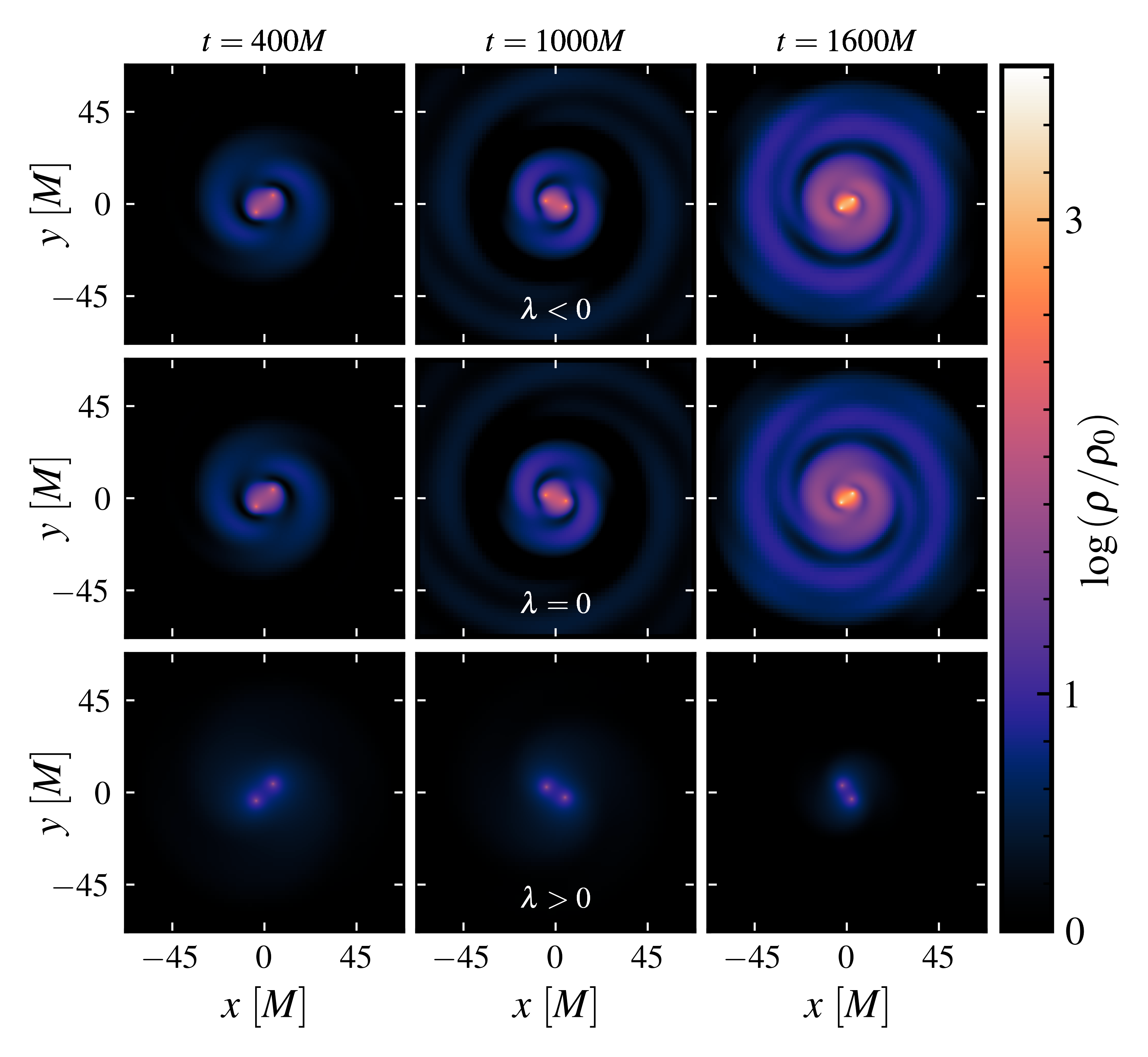}}
    \caption{Evolution of the dark matter cloud for $\lambda=-10^{-4}$ (top), $\lambda=0$ (middle) and $\lambda=0.025$ (bottom). We plot 2D slices of the density. We see that in the repulsive case the overdensity around the binary is suppressed. In the attractive case shown here, the density does not become high enough to trigger a bosenova before the merger occurs, so its evolution is similar to the purely massive case, with only a slight enhancement in the density. A movie of the simulations can be found in \cite{movie}.}
    \label{fig:panel_cloud}
\end{figure}

\begin{figure}[t!]
    \centering
    \includegraphics[width=\linewidth]{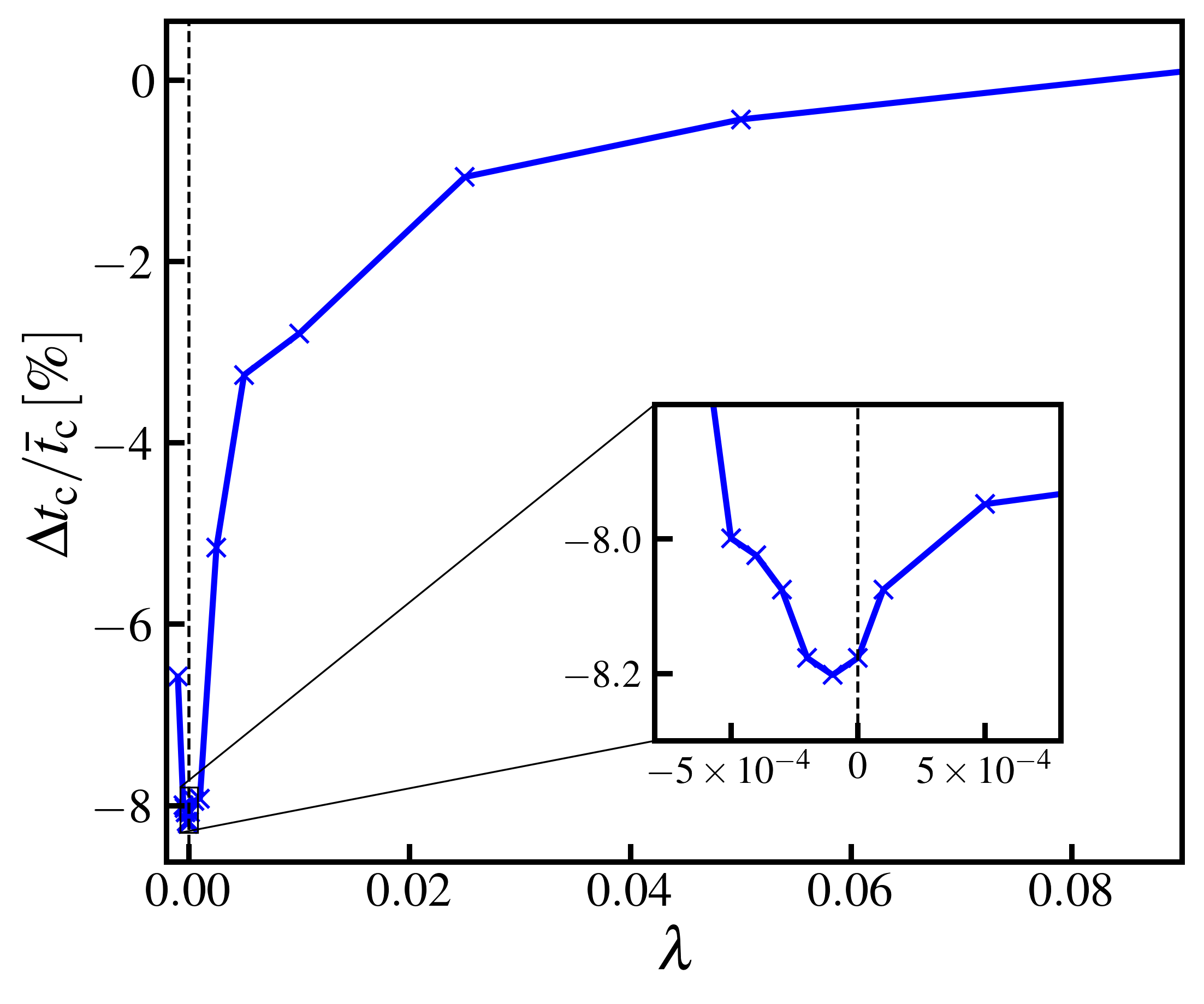}
    \caption{Decrease in the coalescence time $\dtc \equiv t_\mathrm{c} - \bar{t}_\mathrm{c}$ extracted from the gravitational-wave signal, for different self-interaction scenarios. For positive values of $\lambda$ (repulsive interactions), the dephasing is rapidly lost as the self-interaction increases due to the saturation of the cloud density. For small negative $\lambda$ (attractive interactions), the density of the cloud is enhanced and the binary experiences a larger dephasing. However, for sufficiently negative $\lambda$, the bosenova-like event decreases the density and the dephasing is again reduced.}
    \label{fig:dephasing}
\end{figure}

In Fig. \ref{fig:panel_cloud}, we plot the evolution of the density (left to right) for three different self-interactions. The top and middle panels show the evolution of the DM cloud for attractive ($\lambda=-10^{-4}$) and without self-interactions ($\lambda=0$). The profiles remain qualitatively unchanged, with both forming a similar overdensity between the BHs. In contrast, for repulsive self-interactions ($\lambda=0.025$, bottom panel), the density of the DM cloud saturates, and there are significant changes in the structure of the cloud. Unlike the other two cases, repulsive self-interactions suppress the formation of the central overdensity. In Fig. \ref{fig:dephasing}, we plot the relative dephasing of the gravitational-wave signal. Repulsive self-interactions saturate the growth of the cloud, resulting in smaller dephasings. For attractive (but small) self-interactions, the dephasing is slightly larger due to the increased density of the cloud.

We can quantify more precisely the cause of this dephasing by computing the charges and fluxes associated with the currents in the time direction $\xi_t^\nu = (1,0,0,0)$, angular direction $\xi_\phi^\nu = (0,-y,x,0)$, and radial direction $\xi_r^\nu = -(0,x,y,z)/r$. The expressions for the time direction are in Eqns. (\ref{eq:Qt}-\ref{eq:Ft}), and the angular and radial directions
 \begin{align}
    Q_{\lbrace{\phi,r\rbrace}} &= S_i \xi^i_{\lbrace{\phi,r\rbrace}}\,, \\
    F_{\lbrace{\phi,r\rbrace}} &= -N_i \beta^i S_j \xi^j_{\lbrace{\phi,r\rbrace}} + \alpha N_i S^i_j \xi^j_{\lbrace{\phi,r\rbrace}} \,,
\end{align}
where again $N_i = s_i / \sqrt{(\gamma^{jk} s_j s_k)}$ and $s_i=(x,y,z)/r$. We also track the flux through two inner surfaces that move together with the BHs, which introduces additional advection terms to the flux $F^\mathrm{BH} = \alpha N_i^\mathrm{BH}J^i - N_i^\mathrm{BH} \beta^i \left( Q -  S/2\right)$.

As these directions do not correspond to Killing vectors, there must be an exchange of the charge between matter and curvature, which can be estimated by computing the \textit{source} terms associated to these currents
\begin{equation}
    S = \alpha T^\mu_\nu \nabla_\mu \xi^\nu\,.
\end{equation}
It is this quantity that corresponds to gravitational forces in the Newtonian limit, and that quantifies the way in which momentum is extracted from the binary by the matter\footnote{We also include in this quantity the accretion of the matter charge into spheres around the BHs, since the volume integral is not defined at the singularity.}. The expressions for the source terms for the energy, angular and linear momentum currents are
\begin{align}
    S_t =& -\rho \partial_t \alpha + S_i \partial_t \beta^i + \frac{\alpha}{2} S^{ij}\partial_t\gamma_{ij} \\
    S_{\lbrace{\phi,r\rbrace}} =& \alpha S^\mu_\nu \partial_\mu \xi^\nu_{\lbrace{\phi,r\rbrace}} + \alpha S^\mu_\nu {}^{(3)}\Gamma^\nu_{\mu\sigma} \xi^\sigma_{\lbrace{\phi,r\rbrace}} \nonumber \\ 
    & - S_\nu \beta^i \partial_i\xi^\nu_{\lbrace{\phi,r\rbrace}} + S_\nu \xi^\mu_{\lbrace{\phi,r\rbrace}} \partial_\mu\beta^\nu - \rho\xi^\mu_{\lbrace{\phi,r\rbrace}} \partial_\mu \alpha\,
\end{align}
where $\partial_t \gamma_{ij}=-2\alpha K_{ij} +D_i\beta_j + D_j \beta_i$, and both $\partial_t\alpha$ and $\partial_t\beta^i$ are given by our moving puncture gauge conditions. The quantities $\partial_\mu\xi^\nu$ are all zero except $\partial_x \xi^y = 1$ and $\partial_y \xi^x = -1$.

\begin{figure}[t!]
    \centering
    \includegraphics[width=\linewidth]{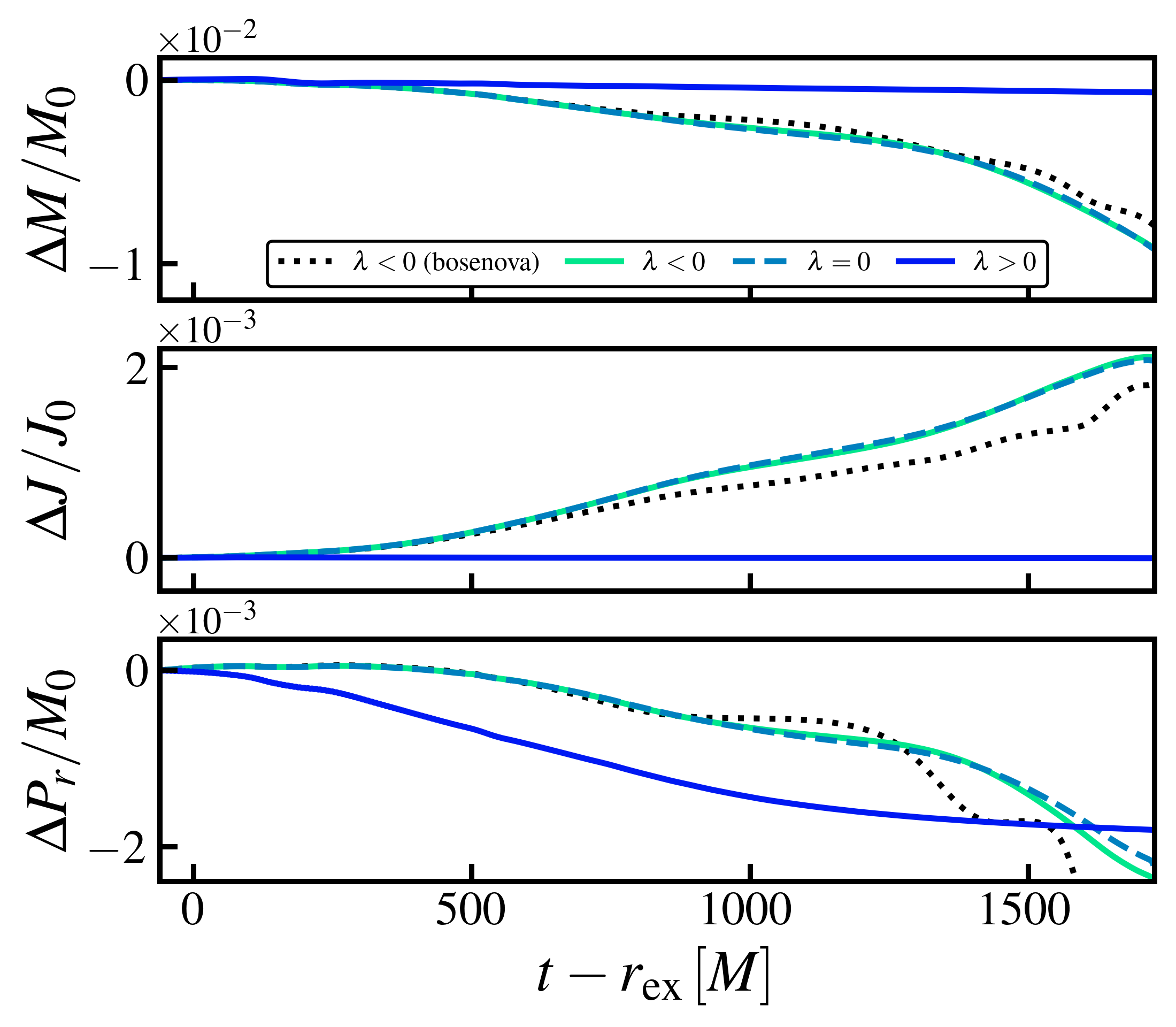}
    \caption{Time integrals of the source terms for the energy, angular and radial momentum currents (top, middle and bottom panels, respectively). These quantify the extraction of mass $\Delta M$, angular momentum $\Delta J$ and radial momentum $\Delta P_r$ from the binary by the matter, illustrating the effect the dark matter cloud has on the binary. We compare the attractive $\lambda=-10^{-4}$ and repulsive $\lambda=0.025$ cases to the $\lambda=0$ case, which are those shown in Fig. \ref{fig:panel_cloud}.}
    \label{fig:fluxes}
\end{figure}

\begin{figure*}[t!]
    \centering
    \href{https://youtu.be/ZeoMlUHknUM}{
    \includegraphics[width=\linewidth]{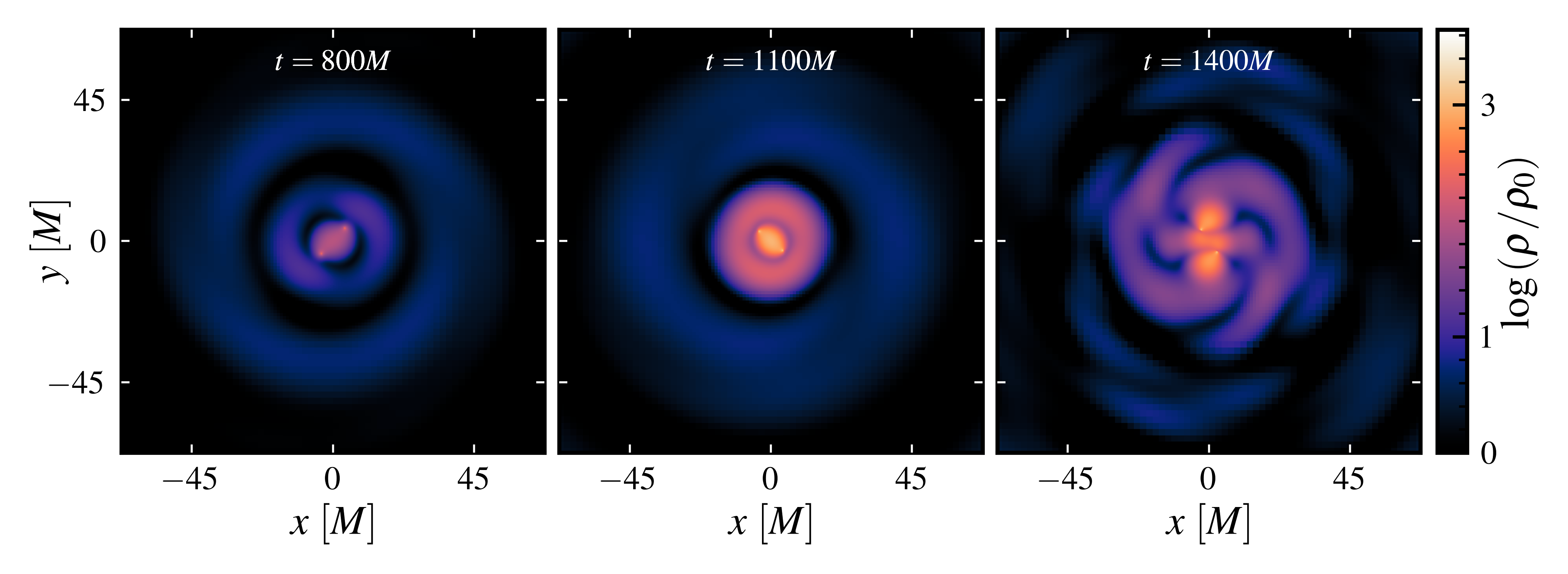}}
    \caption{Bosenova explosion and disruption of the cloud during an equal-mass black hole merger for $\lambda=-10^{-3}$. The density of the dark matter cloud is enhanced due to the attractive self-interactions (compare the middle panel of this figure and those in Fig. \ref{fig:panel_cloud}). However, attractive self-interactions dominate and result in the sudden collapse of the cloud, with a burst of radiation that reduces the cloud density, and hence the dephasing. A movie of the simulation can be found in \cite{movie}.}
    \label{fig:binary_bosenova}
\end{figure*}

In Fig. \ref{fig:fluxes} we plot the time integration of these quantities within a radius of $40M$ from the centre of the binary. These quantify the extraction of mass $\Delta M$, angular momentum $\Delta J$ and radial momentum $\Delta P_r$ from the binary by the matter, which drive the dephasing and the change in the coalescence time $\dtc$ since they affect the balance in the angular momemtum $J=Mr^2\left(2\pi/T\right)$. We compare the attractive $\lambda=-10^{-4}$ and repulsive $\lambda=0.025$ cases to the $\lambda=0$ case. In these cases, the attractive self-interaction gives the greatest relative dephasing at around $8.2\%$, although it is very similar to the purely massive case. We attribute the slight difference in the coalescence time to the slightly higher values of $\Delta J$ and $\Delta P_r$ in Fig. \ref{fig:fluxes}. In the repulsive case, the dephasing is only around $1\%$: the exchanges in energy and angular momentum are highly suppressed (top and middle panels in Fig. \ref{fig:fluxes}), whilst the radial exchange of momentum shown in the bottom panel is smaller.

In principle one may naively expect that increasing the attractive self-interactions further enhances the density of the cloud, but Fig. \ref{fig:dephasing} shows that for $\lambda \lessapprox -10^{-4}$, the dephasing $\dtc$ decreases. This is because, as in the isolated BH case, when the field reaches sufficiently high values, the non-linear attractive interactions of the scalar field dominate and lead to a bosenova-like event. There is a sudden collapse at the overdensity between the binary that results in a burst of matter, disrupting the cloud, as illustrated in Fig. \ref{fig:binary_bosenova}. This behaviour seems to be generic for any kind of attractive self-interaction with $\lambda \vert\Phi\vert^4$ at leading order, but we expect that the exact form of the ejection and recovery to the steady growth phase after the bosenova will depend somewhat on the form of the potential beyond the $\lambda \vert\Phi\vert^4$ regime. The critical value of $\lambda$ at which it occurs will also depend on factors like the asymptotic density, the scalar mass and the parameters of the binary.

\section{Discussion}\label{sec:discussion}

We have studied the evolution of self-interacting dark matter around black holes using numerical simulations. Motivated by EMRIs, we have explored the accretion-driven growth of a scalar DM overdensity around an isolated Schwarzschild BH. We have found that the profile of the cloud is cored for repulsive self-interactions, and more cuspy when these are attractive. Repulsive self-interactions saturate the density of the cloud, while attractive self-interactions lead to a faster growth. However, when attractive self-interactions dominate the dynamics, this enhancement can be temporarily reversed by a bosenova-like event, as the cloud collapses and the system emits a violent burst of energy.

We used numerical relativity simulations that include backreaction onto the metric to study the growth of a DM cloud during an equal-mass BH merger. We have extracted the gravitational waves radiated during the inspiral, merger and ringdown phases and quantified the impact that self-interactions have in the waveform by computing the dephasing of the gravitational-wave signal. We have found that repulsive self-interactions reduce the amount of dephasing compared to the purely massive case because they suppress the DM density that accumulates around the binary. Attractive self-interactions enhance it, but if the cloud grows significantly, the binary triggers a bosenova-like event as in the isolated BH case. The cloud is then disrupted by the motion of the BHs, and the dephasing is again reduced relative to the purely massive case.

In our simulations, we are constrained to use larger values of the density in order to see the dephasing effects on the short timescales on which we are able to simulate the merger. Smaller values should result in similar effects but on longer timescales. To put the values we use into context -- constraints exist on the total mass of dark matter around the supermassive BH at the centre of our galaxy, putting it at less than 0.1\% of the mass of the central BH within $10^4$ Schwarzschild radii using constraints from the S2 star \cite{GRAVITY:2021xju,GRAVITY:2023cjt}. This corresponds to densities of roughly $10^{-15}M^{-2}$ in our units, which is significantly lower than the value of $10^{-9}M^{-2}$ that we use as the asymptotic density. However, the densities we use are lower than possible densities from superradiant clouds where 4-10\% of the BH mass can be contained within a cloud out to around 1-10 Schwarzschild radii \cite{Herdeiro:2021znw}, which gives values up to $\sim 10^{-5}M^{-2}$. They are comparable to those expected from the isolated adiabatic evolution of dark matter spikes \cite{Bertone:2024wbn}.

Whilst the densities that we study are large compared to galactic average values, the self-interactions that we study here are relatively small and not at a level that conflict with current observational constraints. 
The dimensionless coupling for the self-interaction $\bar\lambda$ for a scalar mass $m=\hbar\mu/c$ is constrained by BBN and structure formation to be \cite{Li:2013nal,Arbey:2003sj,Diez-Tejedor:2014naa}
\begin{equation}
    \bar{\lambda} \lesssim 1 \left(\frac{m}{{\rm eV}}\right)^4 ~.
\end{equation}
In our simulations we have fixed the scalar mass to be
$m \approx 5\times 10^{-17}\, \left(M/10^6 M_\odot\right)^{-1}  \,\mathrm{eV}$,
which corresponds to the constraint
\begin{equation}
    \bar{\lambda} \lesssim 6 \times 10^{-66} \left(\frac{M}{ 10^6 M_\odot}\right)^{-4} ~.
\end{equation}
We have chosen the parameter $\lambda\approx \mathcal{O}(0-0.1)M^{-2}$, for which the upper end corresponds to the dimensionless value 
\begin{equation}
    \bar{\lambda} \sim 10^{-89} \left(\frac{M}{ 10^6 M_\odot}\right)^{-2} ~,
\end{equation}
or equivalently a self-interaction energy scale $f = \sqrt{\lambda} / m \sim M_\mathrm{Pl}$. This interaction strength is below the bound above for all realistic BH masses\footnote{For comparison, the QCD axion has $\bar{\lambda} \sim 10^{-53}$ for an axion decay constant $f\sim 10^{12}$ GeV and a mass $m\sim 10^{-5}$ eV \cite{Sikivie:2009qn}.}.

Based on our results, self-interactions appear to mainly reduce rather than enhance dephasing effects. Even for the relatively small values of self-interaction studied, the non-linear part of the potential played a significant role, and both attractive and repulsive values mostly resulted in lower dephasing than the purely massive case. We are only considering a short period close to merger, and so given the much longer timescales available for build-up of the cloud prior to this point, the effects could become relevant well in advance of the actual merger. In the case of attractive self-interactions, this could mean bosenova-like explosions reccuring during the inspiral. The absence or observation of such bursts can be used to put constraints on the existence of attractive scalar DM, as has been proposed in \cite{Budker:2023sex,Arakawa:2024lqr,Arakawa:2023gyq,Eby:2024mhd}. 

As this article was under review, a relevant paper \cite{Takahashi:2024fyq} studied the effect of self-interactions in the context of superradiance.

\section{Acknowledgements}

We thank Joshua Eby, Ben Elder, Rodrigo Vicente and Miguel Zumalacarregui for useful comments and discussions, and the GRChombo collaboration (\href{www.grchombo.org}{www.grchombo.org}) for their support and code development work. JCA acknowledges funding from the Beecroft Trust and The Queen’s College via an extraordinary Junior Research Fellowship (eJRF). JM acknowledges funding from STFC. KC acknowledges funding from the UKRI Ernest Rutherford Fellowship (grant number ST/V003240/1) and STFC Research Grant ST/X000931/1 (Astronomy at Queen Mary 2023-2026). PGF acknowledges support from STFC and the Beecroft Trust. 

This work used the DiRAC Memory Intensive services Cosma8 and Cosma7 at Durham University, managed by the Institute for Computational Cosmology on behalf of the STFC DiRAC HPC Facility (www.dirac.ac.uk). The DiRAC service at Durham was funded by BEIS, UKRI and STFC capital funding, Durham University and STFC operations grants. DiRAC is part of the UKRI Digital Research Infrastructure.

\bibliography{refs.bib}

\end{document}